# The Quasi cellular nets-based models of transport and logistic systems

## Abstract

There are many systems in different subjects such as industry, medicine, transport, social and others, can be discribed on their dynamic of flows. Nowadays models of flows consist of micro- and macro-models. In practice there is a problem of convertation from different levels of simulation.

In the different articles author descriptes quasi cellular nets. Quasi cellular nets are new type of discrete structures without signature. It may be used for simulation instruments. This structures can simulate flows on micro- and macro levels on the single model structure.

In this article described using quasi cellular nets in transport and logistics of open-cast mining.

**Keywords**

quasi cellular nets, model, simulation, logistics, transport, open-cast mining logistics.

**Author**

Aristov Anton Olegovich, Russia, Moscow, National University of Science and Technology "MISIS", docent, candidate of technical sciences (computer science), 119991, Russia, Moscow, Leninsky pr. 6, 8-906-786-7124, batan-87@mail.ru

# Concepts

**Term 1.** Static structure of 2D (3D) quasi cellular net (fig. 1) if discrete structure consisted of multitude $Q=\{Q_1,Q_2,...,Q_n\}$ of circular elements in 2D (3D) coordinates. Each elements have constant radius $R$ and weights from multitudes $X=\{x_1,x_2,...,x_n\}$, $Y=\{y_1,y_2,...,y_n\}$. For each $Q_u \in Q$ at least one $Q_v \in Q$ ( $u,v=1,2,...,n; u \neq v$ ) that $(x_u-x_v)^2+(y_u-y_v)^2 \leq 4R^2$.

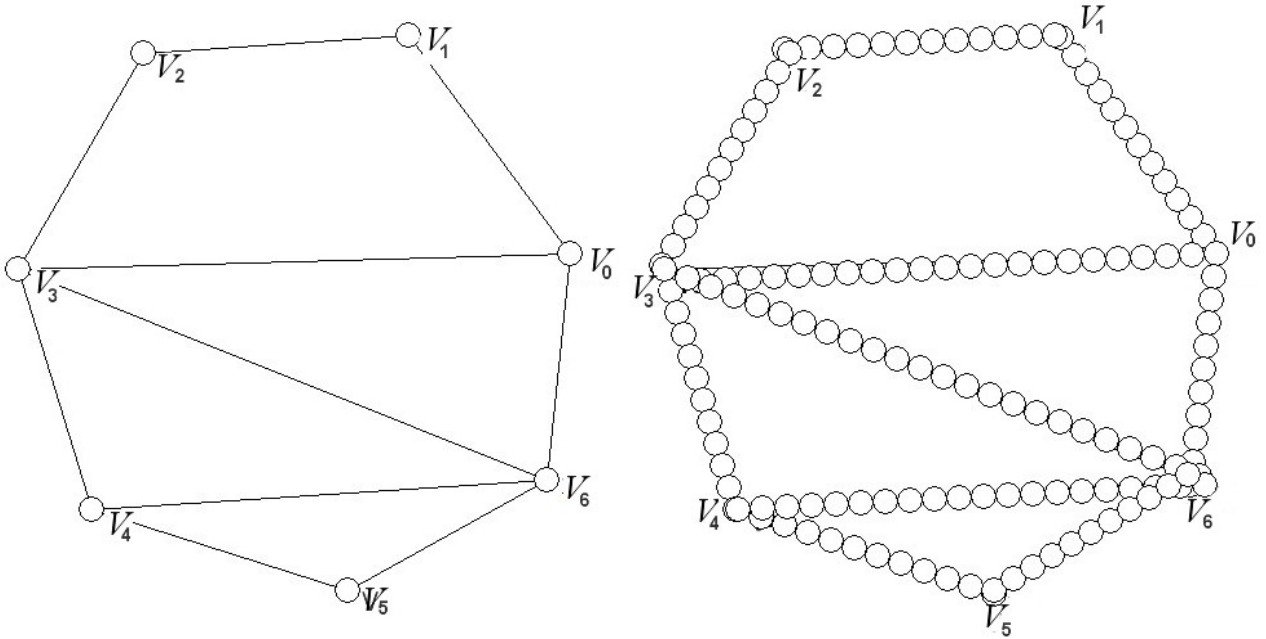

*Figure 1. Static structure of quasi cellular net (based on graph model)*

**Term 2.** Each element of $Q=\{Q_1,Q_2,...,Q_n\}$ (see term 1) this is *element of quasi cellular net*. It also called *"cell"*.

**Term 3.** State of quasi cellular net is variable discrete or continuous value that weight each cell of quasi cellular net.

**Term 4.** Structure of cell of quasi cellular net is a set of variables and constants their weight each cell $Q_i=(B_i,C_i,S_i)$, where $B_i=(B_1,B_2,...)$ – constant (basic) cell parameters, $C_i=(C_1,C_2,...)$ – changeable cell parameters, $S_i=(S_1,S_2,...)$ – parameters of micro object that located in cell. It also called "state variables" (phase variables).

**Term 5.** Condition of neighborhood between $Q_u=(B_{u1},B_{u2},...,C_u,S_u)$ and $Q_v=(B_{v1},B_{v2},...,C_v,S_v)$ – condition, defined on *neighborhood predicate*:
$$P(Q_u,Q_v)=P(B_{u1},B_{u2},...,B_{v1},B_{v2},...)=\begin{cases} 0, npu \neg f(B_{u1},B_{u2},...,B_{v1},B_{v2},...) \\ 1, npu\ f(B_{u1},B_{u2},...,B_{v1},B_{v2},...) \end{cases},$$
где $f(B_{u1},B_{u2},...,B_{v1},B_{v2},...)$ – expression, that defined condition of state transition between cells. For 2D quasi cellular net elements $Q_u=(x_u,y_u,C_u,S_u)$ and $Q_v=(x_v,y_v,C_v,S_v)$ :
$$P(Q_u,Q_v)=P(x_u,y_u,x_v,y_v)=\begin{cases} 0, npu (x_u-x_v)^2+(y_u-y_v)^2 > 4R^2 \\ 1, npu (x_u-x_v)^2+(y_u-y_v)^2 \leq 4R^2 \end{cases}$$
For 3D quasi cellular net elements $Q_u=(x_u,y_u,z_u,C_u,S_u)$ and $Q_v=(x_v,y_v,z_v,C_v,S_v)$ :

$$P(Q_u, Q_v) = P(x_u, y_u, z_u, x_v, y_v, z_v) = \begin{cases} 0, npu\ (x_u - x_v)^2 + (y_u - y_v)^2 + (z_u - z_v)^2 > 4R^2 \\ 1, npu\ (x_u - x_v)^2 + (y_u - y_v)^2 + (z_u - z_v)^2 \leq 4R^2 \end{cases}$$

**Term 6.** Simulation time $t = 0, \theta, 2\theta, 3\theta, ...$, where $\theta$ - constant timestap. In each timestap state of each cells of quasi cellular nets can be change.

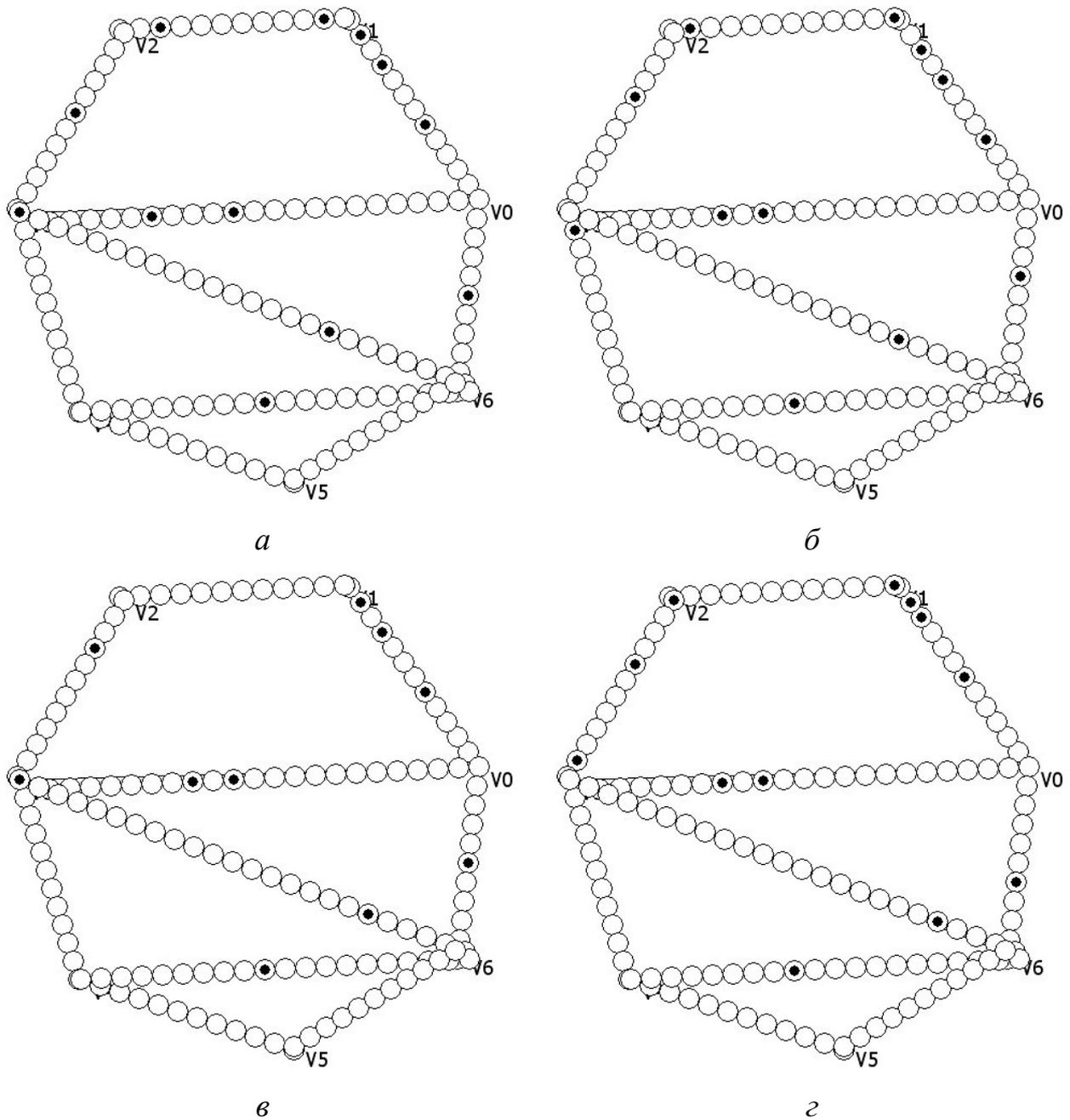

*Figure 2. Circulation in quasi cellular net with binary state.*
*State of quasi-cellular net: a — at the moment of simulation time t,*
*b — at the moment of simulation time $t+\theta$, c — at the moment of simulation time $t+2\theta$,*
*d — at the moment of simulation time $t+3\theta$*

**Term 7.** Transition of discrete state $Q_u \rightarrow Q_v$ in quasi cellular net is change of cell state

$$\begin{cases} S_u(t)=S \\ P(Q_u,Q_v)=1 \end{cases} \rightarrow \begin{cases} S_v(t+\theta)=S_u(t) \\ S_u(t+\theta)=\overline{S} \end{cases}$$, where $S$ – "transitable" (flow-based) state, $\overline{S}$ – non-transitable state that set in cell after "transitable" state ($S$) transition to neighbor cell.

**Term 8.** Transition of continuous state $Q_u \rightarrow Q_v$ кin quasi cellular net is change of cell state $P(Q_u,Q_v)=1 \rightarrow \begin{cases} S_v(t+\theta)=S_v(t)+\delta \\ S_u(t+\theta)=S_u(t)-\delta \end{cases}$, where $\delta$ – value of state difference.

**Term 9.** Circulation on quasi cellular net with discrete state is multitude of transitions (see term 7) $\{Q_u \rightarrow Q_v \mid S_u(t)=S, P(Q_u,Q_v)=1\}$ on each timestap of simulation time $t=0,\theta,2\theta,3\theta,...$ .

**Term 10.** Circulation on quasi cellular net with continuous state is multitude of transitions (see term 7) on $\{Q_u \rightarrow Q_v \mid P(Q_u,Q_v)=1\}$ each timestap of simulation time $t=0,\theta,2\theta,3\theta,...$ . For example – circulation in quasi cellular net with binary state (fig. 2).

**Term 11.** Constraint condition of quasi cellular nets is condition that constraint execution of transitions (see terms 7,8).

**Term 12.** Generator (inflow) in quasi cellular nets is cell $Q_g=(B_g,C_g,S_g(t))$, that $S_g(t+\theta)=S_g(t)+f(t)$, where $f(t)$ – discrete or continuous function of generation.

**Term 13.** Outflow in quasi cellular net is cell $Q_o=(B_o,C_o,S_o(t))$ that $S_o(t+\theta)=S_0$ independent of circulation, where $S_0$ – "clear" ("zero") cell state.

**Term 14.** Turnstile (cell of delay) in quasi cellular net is cell that save state $S_k$ till the time $\tau$.
$Q^{(T)}=(x_T,y_T,S_T(t),t_m+\tau,O^{(T)}(t))$, $S_T(t)\in\{S_k,\overline{S_k}\}$

$$\begin{cases} S_T(t_m)=S_k \\ O^T(t_m)=1 \end{cases} \rightarrow \begin{cases} O^{(T)}(t_m)=0 \\ S_T(t)=S_k \mid t_m<t<t_m+\tau \\ O^{(T)}(t_m+\tau)=1 \end{cases},$$

where $S_k$ – saved state, $\tau$ – saving time; $\overline{S_k}$ – state of open turnstile and state for transition turnstile to $S_k$; $t_m$ - time moment when turnstile was closed after transition to $S_k$; $O^{(T)}$ – internal state of turnstile (open/closed) $O^{(T)}\in\{0,1\}$.

## Transport systems

In other articles of author considered using quasi cellular nets in public spaces. Coordinate quasi cellular nets may be used for simulations of flows of people. There is analogy between transport and people flows. Let's describe aspects of using quasi cellular nets in transport and logistics.

In analogy between transport and people flows, flows on logistics and transportt based on dynamics of transitable micro- objects and their moving on space. In this sitation cars are elements of this flow on micro- level. In logistic systems there are several types of flows, such as transports and other material resources, finance or informational flows. These flows move in space that's why is advisable to use coordinates quasi cellular nets.

For creation model of logistical system based on quasi cellular nets requires to define cell structure, interpret elements of quasi cellular nets (cells, inflows, outflows, turnstile) and choice type of circulation.

Often geometry of roads defined as graph models, where vertexes are simulate intersections and edges are simulate straight roads[1,2]. In this model also use algorythm of Ford and Falkerson[3]. In difference of this models quasi cellular nets may be also used for simulations in micro-levels. That's why is advisable to use method of basic graph to sththesis quasi cellular nets for transport systems. The model of basic graph and quasi cellular nets (fig. 3). Size of each cell is matches to size of car and other transport.

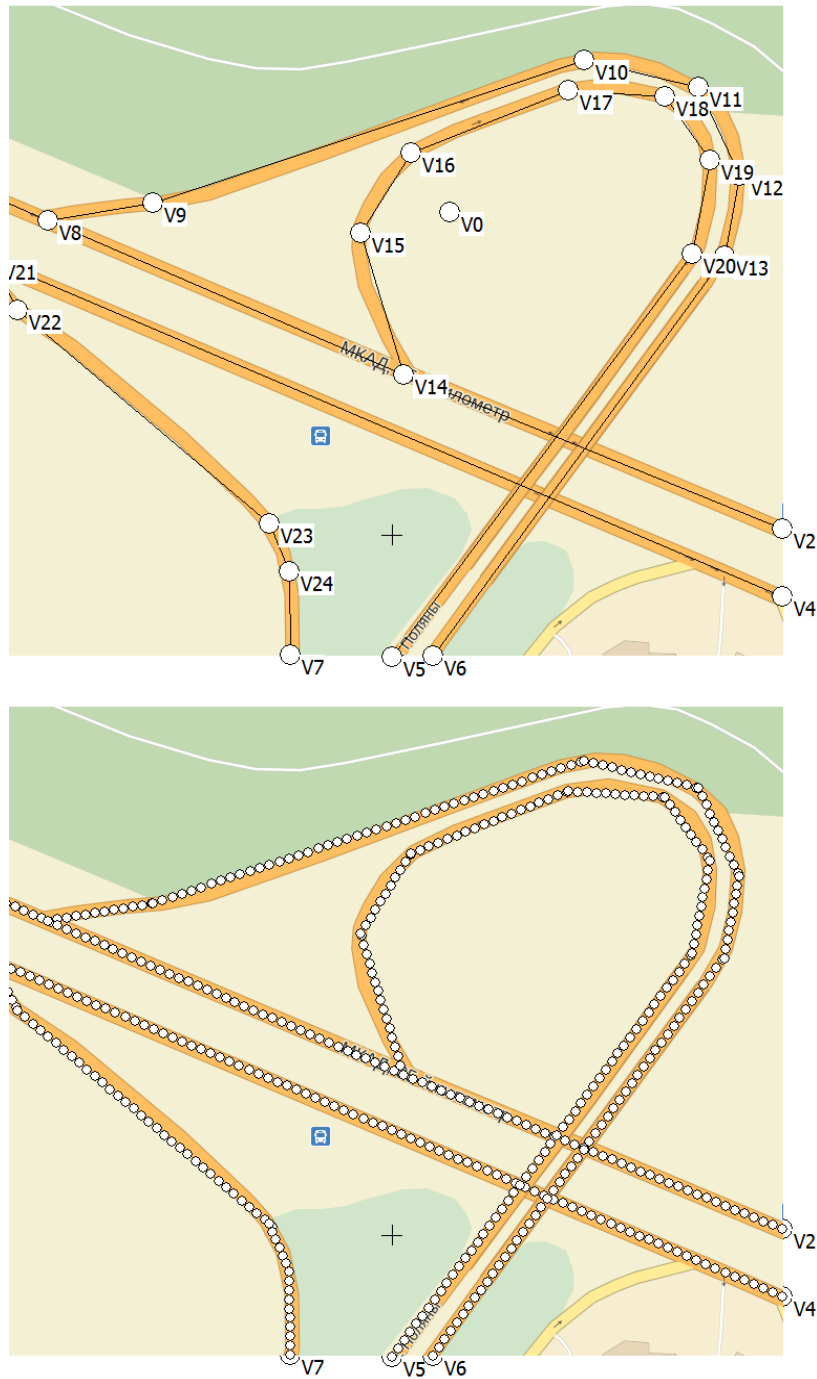

*Figure 3. Method of basic graph and quasi cellular net*

It also can should pay attention that in this model use only plank road section. For create quasi cellular model of multiband plank road section may be used order (fig. 4).

In this situation may be elements of cellular automata, but if there is stripe separator then cells $Q_u$ и $Q_v$ are not satisfy the condition:

$$(x_u - x_v)^2 + (y_u - y_v)^2 \leq 4 \cdot R^2 \quad , \qquad (1)$$

где $x_u$, $y_u$, $x_v$, $y_v$.

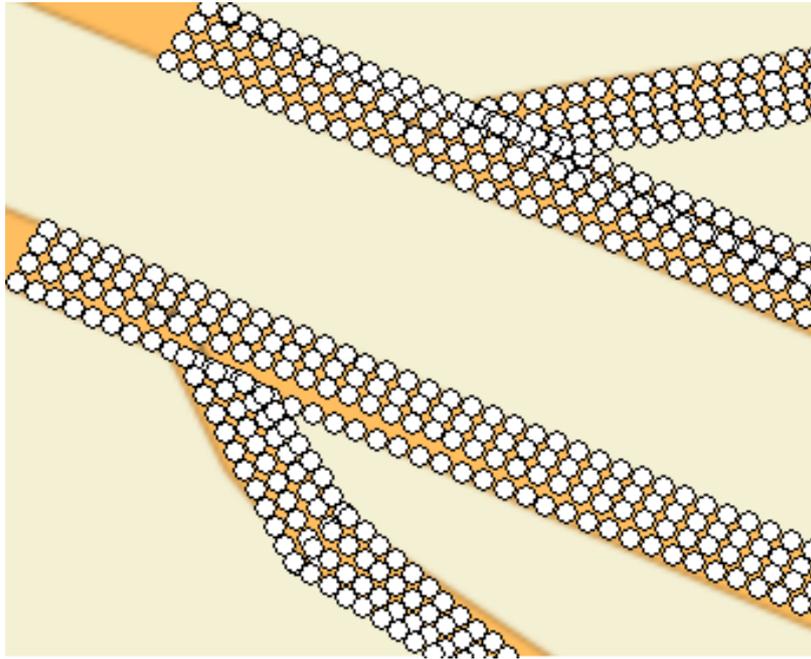

*Figure 4. Structure of quasi cellular net for multiband plank road*

Dynamics aspects define by circulation of quasi cellular nets. With modern models of cars and other transports [1,2] and features of their simulation on micro- level[2] appropriate to use microcirculation in quasi cellular nets. That's why cell structure is[4]:

$$Q_u = (x_u, y_u, dx_u, dy_u, S_u) \quad , \qquad (2)$$

where $(dx_u, dy_u)$ – vector of translation to the neighbor cell, $S_u$ – binary value of presence transport in this part of road (cell). Therefore, we have:

$$\begin{cases} x_v = x_u + dx_u \\ y_v = y_u + dy_u \end{cases} , \qquad (3)$$

if it is true (1).

Sometimes (on intersection, rotates, in the model of stripe changes) may be used moving in ither directions. Moving in other direction may be with probability. For this situation in structure $Q_u$ need set vector:

$$(dx_{ui}, dy_{ui}, p_{ui}) \quad , \qquad (4)$$

where $p_{ui}$ – probability of translation from cell $Q_u$ to direction $(dx_{ui}, dy_{ui})$.

So it were described aspects of creations quasi-cellular nets for transport flows.

## Models of quarry logistics

Next ideas of quasi cellular nets usings is subject interpret their elements. For example, turnstile can simulate differential motion. In some systems there are physics analogies of turnstile such as terminals of payd roads and crosswalks.

It also can should pay attention that is analogies between transport logistics and other types of logistics such as industry. Porcesses in logistics dependency of transport and resources between different stages of processing and transformations of this resources. Therefore for models of logistical systems based on quasi-cellular nets, requires define subjects interpretations of turnstile and transformers.

For example consider simulation of transport logistics of sandpit. Trajectories of sandpit

transports (fig.5)

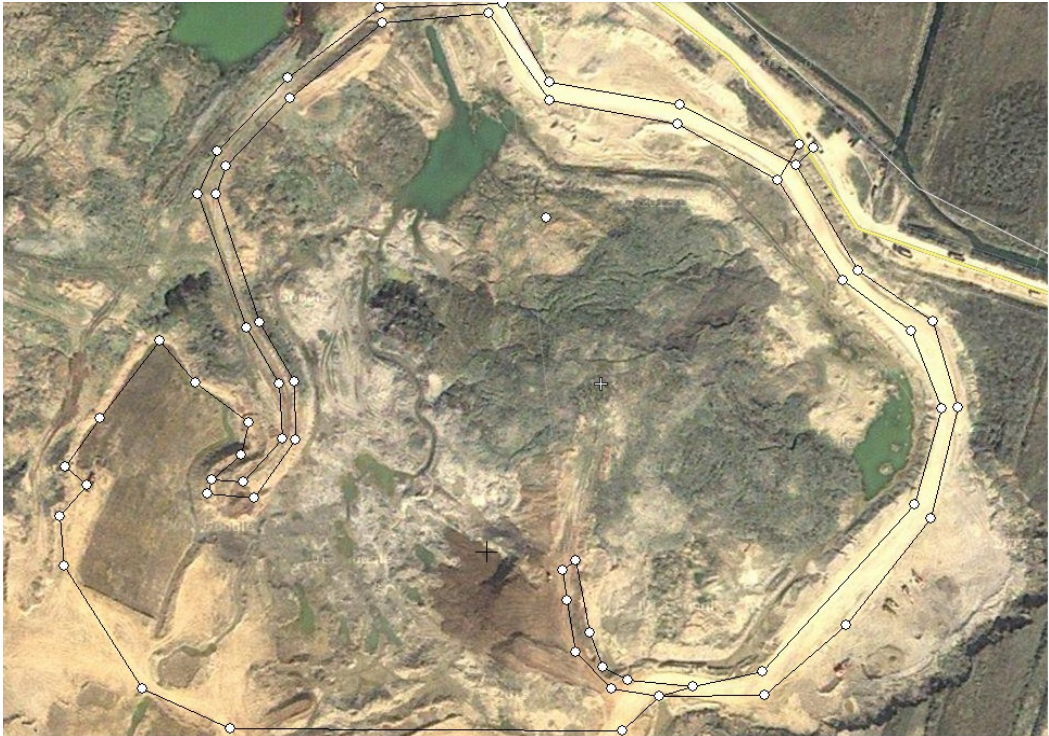

*Figure 5. Trajectories of sandpit transports*

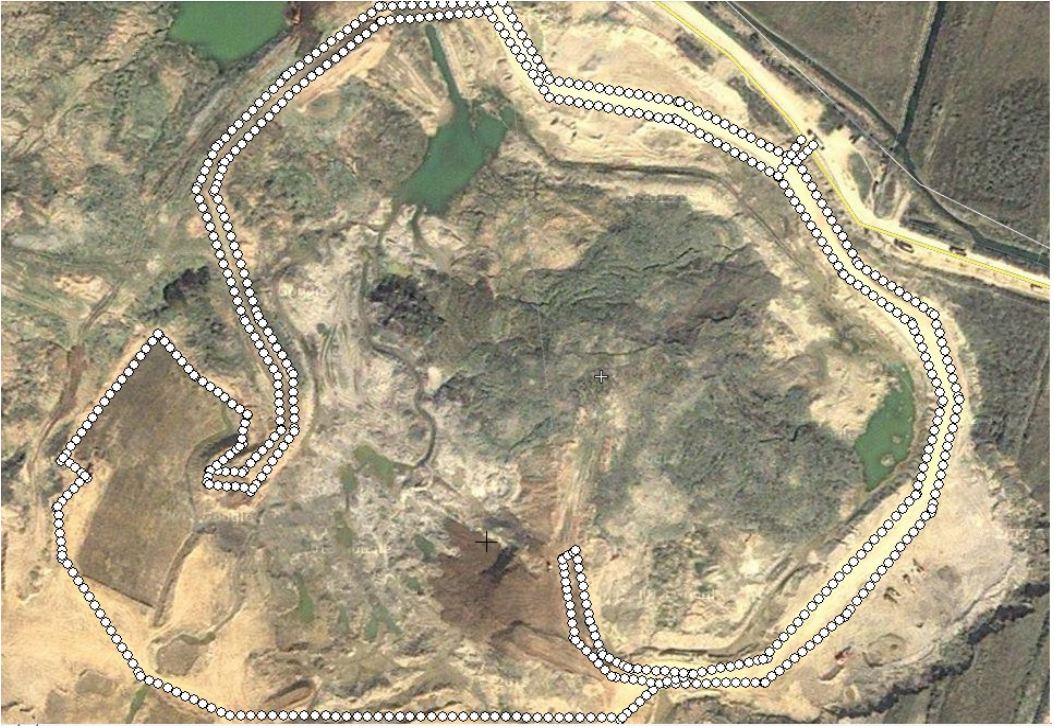

*Figure 6. Quasi cellular net for sandpit simulation*

The main operations of sandpit transport logistics are loading and transportation of sand. That's why basic graph (fig.5) and quasi-cellular nets (fig.6) should be considered as scheme of freight and each part of paths is cell. In this cells are tippers these move on trajectories. That's why structure of cell is (2). Other parameters of cell are microparameters of tipper, such as volume transported rock mass (sand), fuel consumption, the emission of exhaust gases, and so on. n. It also use micro circulation, because it is moving on trajectories (3).

It also should pay attention to inflow and outflow. The one of the most actual problem of logistical systems are borders of this of this systems. For example, if should be simulate full cycle of loading and ore transportation to processing plant and moving of empty tippers then inflow and outflow are not using. If required only model of quarry then inflows simulate entrance to the quarry territory and outflows simulate exit from this territory. Generally structure of cell is dependent of features of models and other conditions these supplement (2) .

For example turnstile. If state of each cell is availability of tipper in thiis part of space then save of this state for a long time is loading process. In this formulation, turnstile simulate excavator that loading tippers. So this model may be used for evaluation mining systems.

It also need describe polysort flows in logisticmodels. For example raw materials, semi-finished products, finished products, packaged products etc. In mining transport of sandpit polysort flows defined as composition of sand (clear sand, water, stones, clay). This values should be include to (2).

In practice of computer simulation there are not definitions between polysort flows on different subjects. But in models of logistical systems the main problem is flow transformations these simulate different processes in logistics, such as manufacturing, processing, packaging, labeling etc.

## Conclusion

So in this article were described common aspects of creation models of logistical systems based om quasi-cellular nets and these interpretations in logistical systems. Practical logistical modeling problems are closely related with evaluation and researching different flow characteristics. But in this article are described are general guidelines of using quasi-cellular nets for logistical models.

## References

1. Buslaev A.P., Lebedev A.A., Yashina M.V. (2011) Flow simulation based on graph models.Theory and mathematics aspects // Moscow : Moscow State Automobile and Road Institute. 105 s.
2. Badalyan A.M., Eremin V.M. (2007) Computer simulations of conflict situations for researching assess the level of traffic safety on two-lane road // Moscow : Katalog —240 s.
3. Kristofides N. (1978) Theory of graphs. Algorithmic approach // Moscow : MIR — 432 s.
4. Aristov A.O. (2014) Theory of quasi cellular nets // Moscow : National University of Science and Technology "MISIS". – 188 s.